\documentstyle[eqsecnum,multicol,aps,psfig,a4]{revtex}
\draft

\headheight=\baselineskip
\begin{document}
 
\bibliographystyle{prsty}

%\narrowtext
%\twocolumn
%\onecolumn

\title{On a certain analogy between hydrodynamic flow in 
porous media and heat conductance in solids}

\author{F. Tzschichholz$^{1,2}$}
 
\address{
$^{1}$ Institut f\"ur Computeranwendungen I, Universit\"at Stuttgart, 
Pfaffenwaldring 27, D-70569 Stuttgart, Germany.
}
\smallskip
\address{
$^{2}$ Department of Physics, Norwegian University of Science and Technology, 
N--7034 Trondheim, Norway}

\date{\today}
\maketitle
 
\begin{abstract}
We consider a porous medium being saturated with a pore fluid (Biot's 
theory). The fluid is assumed as incompressible. It is shown
that the general integral of the elastic and pressure equations
can be written in form of a time dependent vectorpotential 
${\bf F}$ being a solution of a homogeneous, fourth order 
differential equation. The obtained equation for ${\bf F}$ is of a more
general form than the corresponding thermo-elastic
vectorpotential, being a solution of a time dependent and 
inhomogeneous vector bi-Laplacian.  
Both vectorpotentials do, however, agree for stationary problems in 
general and for certain particular boundary conditions (irrotational
deformations). An example of an irrotational deformation is studied
in detail, exhibiting known properties of classical vector diffusion. 
\end{abstract}
 
\pacs{PACS number(s):?? }

\newsavebox{\boxl}
\savebox{\boxl}{$\ell$}
 
\begin{multicols}{2}
\section{Introduction}\label{section:introduction}

Imposing an inhomogeneous pressure/temperature field on a (porous) material 
causes in general mechanical deformations and stresses. 
These `induced' elastic deformations in turn can and do 
influence the further mass/heat transport. 
The problem is described as a coupled field problem consisting
of (at least) two equations (a) an elastic equation describing 
mechanical displacements due to a given source field
(pressure/temperature gradient field) and (b) at least one transport 
equation for mass/heat with a source term originating from the 
mechanical deformation (divergence of the displacement field).

The entire description bases on continuity equations and
constitutive laws which are discussed in textbooks in great
detail\cite{Pagano}. Perhaps less known is Biot's theory of
hydrodynamic flow in porous elastic media\cite{Biot41}. 
We will
address this problem below in more detail.
In particular we wish to draw some attention on similarities for the
flow and the heat problem.

\section{General considerations}\label{section:general_considerations}
We give the basic outline of Biot's theory for the particular case of 
an incompressible pore fluid. This considerably simplifies the
governing equations and their discussion, because all material
derivaties can be eliminated.
First we discuss the basic elastic equations.
This is followed by a
detailed discussion of the flow equation for incompressible pore fluids.
Thereafter we discuss some properties of the coupled equations, and 
finally establish their general integral following the spirit 
of Galerkin's method.
We specialize the obtained general integral for the case of 
irrotational displacement fields and make the connection to the
problem of thermoelasticity.   
\subsection{Preliminaries}\label{subsection:preliminaries}
In the following bold-face letters always indicate vectors/tensors.
\subsubsection{Lam\'e equation}\label{subsubsection:lame_equation}
The continuity equation for momentum density,
\begin{equation}
\label{equation:momentum_continuity}
div\;{\bf \sigma}\; + {\bf f} = \rho_s \ddot{\bf u},
\end{equation}
represents the mechanical equilibrium condition, where 
${\bf \sigma}=(\sigma_{ij})$ is the stress tensor, 
${\bf f}$ any external body
force, $\rho_s$ the solid's mass density and ${\bf u}=(u_i)$ the mechanical
displacement. We will neglect the inertial terms in the following and 
will consider Biot's theory of porous media
for the case of symmetric elasticity, i.e. 
$\sigma_{ij}=\sigma_{ji}$. Deformations are characterized by 
the strains $2u_{ij}=(\partial u_i/\partial x_j + \partial u_j/\partial
u_i)$. Employing Hooke's linear relation between stresses and strains
for isotropic materials one obtains Lam\'e's equation for the 
displacements,
\begin{equation}
\label{equation:lame}
(1-2\nu)\Delta\;{\bf u} + grad\;div\;{\bf u} +
\frac{2}{E}(1+\nu)(1-2\nu)\; {\bf f} = {\bf 0},
\end{equation}
with Poisson contraction number $\nu$ and Young modulus E. 
There are a few quite general results about this equation, which will 
be useful in the following. Taking the divergence of 
Eq.(\ref{equation:lame}) one has,
\begin{equation}
\label{equation:lame1}
\Delta\;div\;{\bf u} = - \frac{(1+\nu)(1-2\nu)}{(1-\nu)E}\;div\;{\bf
f},
\end{equation}
i.e., $div\;{\bf u}$ is the solution of a Poisson equation and for 
vanishing body forces ${\bf f}$ it is even a potential function. The
physical significance of $div\;{\bf u}$ is that it measures the local
relative volume change due to the deformation 
which is an invariant. 
Pure shear deformations 
are unreflected in Eq.(\ref{equation:lame1}) because they preserve the
volume. Of course all information on vorticity has been lost
by passing from Eq.(\ref{equation:lame}) to Eq.(\ref{equation:lame1}).
On the other hand taking the Laplacian on Eq.(\ref{equation:lame}) and
inserting Eq.(\ref{equation:lame1}) into yields,
\begin{equation}
\label{equation:lame2}
\Delta\;\Delta\;{\bf u}= \frac{2(1+\nu)}{E}\Bigl(
\frac{1}{2(1-\nu)}\;grad\;div\;{\bf f} - \Delta\;{\bf f}\Bigr).
\end{equation}
This equation is of fourth order while Eq.(\ref{equation:lame}) is of
second order; hence their solutions will be in general not
identical. In his seminal work Galerkin resolved this problem. 
He introduced
another vector function ${\bf F}$ satisfying the (inhomogeneous)
biharmonic equation,
\begin{equation}
\label{equation:garlekin}
\Delta\;\Delta\;{\bf F}= -\frac{2}{E}(1+\nu)\; {\bf f},
\end{equation}
with ${\bf f}$ being the same body force as in
Eq.(\ref{equation:lame}).
He showed that the integral of Eq.(\ref{equation:lame}) can be
expressed in form of derivatives of ${\bf F}$,
\begin{equation}
\label{equation:garlekin1}
{\bf u}\;= \Delta\;{\bf F} -\frac{1}{2(1-\nu)}\;grad\;div\;{\bf F}.
\end{equation}
We would like to note that the foregoing considerations 
made no particular assumptions on the body force field ${\bf f}$ in 
particular they apply to conservative fields.
Typically ${\bf f}$ is explicitly known like for 
gravitational forces, or it is itself a solution 
of another independent equation. The latter is the case for 
thermo-elastic problems, in which 
\begin{equation}
\label{equation:fourier}
{\bf f}=-\alpha\;K\; grad\;T 
\end{equation}
where $\alpha$ is the heat expansion coefficient, 
the $K=\frac{E}{3(1-2\nu)}$ accounts for the material's compressibility,
and $T$ is the temperature field. 
Typically Eqs.(\ref{equation:lame}) and (\ref{equation:fourier}) 
are supplemented by a heat conductance equation,
\begin{equation}
\label{equation:heat}
\chi \Delta\;T = \frac{\partial}{\partial t}T,
\end{equation}
with $\chi$ being the coefficient of heat conductance divided by 
the specific heat capacity. 
Equation (\ref{equation:garlekin}) then becomes 
a biharmonic equation with 
time dependent --but known inhomogeneitis-- which do {\em not} dependent 
on the deformation themselves.
As will be shown further below does flow in porous 
materials imply in general 
`body forces' being deformation dependent themselves. 
The deformation acts back on the flow. 
We will come back to this point in more detail further below.

\subsubsection{Flow equation}
\label{subsubsection:flow_equation}
Lets consider fluid flow in a porous continuum. The porosity
(local volumetric fluid fraction) can be described by a scalar field 
$\phi ({\bf r}, t)$. It is common to assume that the relative 
fluid-solid velocity $\phi ({\bf v} - \dot{\bf u})$
can be expressed by Darcy's relation,
\begin{equation}
\label{equation:darcy}
{\bf f} = -\;grad\;p = 
\phi\frac{\kappa}{\mu}({\bf v} - \dot{\bf u}),
\end{equation}
where ${\bf v}$ and $\dot{\bf u}$ are the fluid and solid velocities, 
$\kappa$ the mechanical permeability, $\mu$ the fluid viscosity, and 
$p$ the hydrostatic pressure field.

The continuity equations for solid and fluid mass are,
\begin{equation}
\label{equation:solid_continuity}
\frac{\partial}{\partial t}\Bigl(
(1-\phi)\rho_s\Bigr) + div\;\Bigl(
(1-\phi)\rho_s \dot{\bf u}\Bigr) = 0,
\end{equation}
and
\begin{equation}
\label{equation:fluid_continuity}
\frac{\partial}{\partial t}\Bigl(
\phi\rho_l\Bigr) + div\;\Bigl(
\phi\rho_l {\bf v}\Bigr) = 0,
\end{equation}
respectively, where $\rho_s$ and $\rho_l$ denote the solid and fluid
mass densities.
In the following we make the two important assumptions that the 
mass densities are constants ($\rho_s =const.$ and $\rho_l =const.$). 

%In the limiting case of zero porosity this implies from 
%Eq.(\ref{equation:solid_continuity}) 
%$div\;\dot{\bf u}=0$, which is correct because the elastic solution we
%seek in that case is time independent (zero inertial terms).
%On the other hand for $\phi =1$ we have fluid only; 
%fluid mass conservation results in $div\;{\bf v}=0$. Therefore 
%the isolated fluid phase is assumed to be incompressible.  
 
From Eqs.(\ref{equation:darcy}), (\ref{equation:solid_continuity}) and 
(\ref{equation:fluid_continuity}) and the assumption of constant mass
densities follows a) that the local volume is conserved because the 
inflow of fluid volume into a volume element equals the
solid volume outflow, and b) that the flow equation 
takes the following simple form ($\kappa/\mu =const$),
\begin{equation}
\label{equation:flow}
\frac{\kappa}{\mu}\;\Delta\;p =\; div\;\dot{\bf u},
\end{equation}
i.e. a Poisson equation for the pressure with the rate of the relative
elastic volume change as source term. 
The porosity $\phi$ does not enter explicitly into the flow
equation because of constant mass densities. As soon as $\dot{\bf u}$ 
has been determined from the coupled Eqs.(\ref{equation:lame}) and 
(\ref{equation:flow}) $\phi$ can be computed from 
Eq.(\ref{equation:solid_continuity}).

\subsection{The coupled system}
\label{subsection:the_coupled_system}
\subsubsection{The Galerkin-Biot potential}
\label{subsubsection:galerkin_biot_potential}

After the above preparations we will focus on the coupled 
Eqs.(\ref{equation:lame}) and (\ref{equation:flow}) describing 
elastic--flow interactions for incompressible pore fluids. Note that 
\begin{equation}
\label{equation:coupling}
{\bf f}\;=-\;grad\;p, 
\end{equation}
in Eq.(\ref{equation:lame}), comp. 
Eq.(\ref{equation:darcy}).
In general Lam\'e's vector equation (\ref{equation:lame})
is much harder to solve than 
the scalar Poisson Equation (\ref{equation:flow}). However,
for the {\em very} particular case of irrotational displacements 
Lam\'e's equation can be integrated easily, see Sec.
\ref{subsection:pressures_and_displacements}.
In general (and also for most boundary conditions of practical interest)
displacements and velocities will have a non-zero rotational part. 
Therefore $div\;{\bf u}$ or its time derivative will not contain the
full information about displacements and strains, i.e. all elastic 
volume preserving deformations (shears) are unreflected in it.

%One can of course try to start with Eq.(\ref{equation:lame1}) instead 
%of Lam\'e's equation. This combined with the flow equation 
%results in a heat conductance equation for $div\;{\bf u}$ which could
%be even with appropriate boundary and initial conditions (in 
%$div\;{\bf u}$). Once one knows $div\;{\bf u}$ one certainly knows
%$grad\;div\;{\bf u}$, but $grad\;p$ still needs to be calculated and hence 
%Eq.(\ref{equation:lame}) reduces to a vector Poisson equation in 
%${\bf u}$ (or more precisely $curl\;curl\;{\bf u}= {\bf A}({\bf r},t)$
%with known ${\bf A}$) which needs to be supplemented with boundary 
%conditions compatible to those for the earlier $div\;{\bf u}$ 
%integration.

It has been convincingly demonstrated that for heat-conductance
in solids the relative volume changes,  
$div\;\dot{\bf u}$,   
actually do also appear in Eq.(\ref{equation:heat}). 
However, the corresponding
coefficient is for most solids so small that this term can be safely
disregarded simplifying the theoretical treatment
considerable\cite{Landau65}.  
As an analogy to heat-conductance in solids 
one can think of a {\em strongly} compressible 
fluid, i.e. a gas in a porous medium. One can neglect the back-influence 
of the pressure induced elastic distortion on the pore pressure field 
and is basically left with the equations decribing 
a corresponding thermo-elastic problem.  
On the other hand does flow of an {\em nearly incompressible} fluid 
in a porous medium represent the `opposite' case: 
in order flow to happen 
elastic deformation must take place. Without elastic deformation there is no
flow and vice versa. 

We reconsider the flow problem Eq.(\ref{equation:garlekin}) and make 
some small refinements. 
Of course the pore pressure field is unknown in
Eq.(\ref{equation:garlekin}) since we have not incorporated any 
information from Eq.(\ref{equation:flow}). This needs to be done 
in order to obtain a fully decoupled equation for displacements.
Starting from Eq.(\ref{equation:flow}) we have,
\begin{equation}
\label{equation:poisson}
{\bf f}\;=-\;grad\;p=\frac{\mu}{\kappa}\;grad\;\int_{V'}
\frac{div\;\dot{\bf u}({\bf r'})}
{4\pi \vert {\bf r}-{\bf r'}\vert }\;dV'.
\end{equation}  
This equation still needs to be complemented by boundary conditions
for the pressure field.
Inserting Eq.(\ref{equation:garlekin1}) into 
Eq.(\ref{equation:poisson}) and integrating employing Helmholtz's 
theorem\cite{Morse53} 
one obtains a relation between $\bf f$ and $\bf F$ which
is in turn inserted in Eq.(\ref{equation:garlekin}),
\begin{equation}
\label{equation:garlekin2}
\Delta\;\Delta\;{\bf F}= 
\frac{1}{D}\; 
grad\;div\;\dot{\bf F},
\end{equation}
with the abbreviation 
\begin{equation}
D=\frac{\kappa}{\mu}\frac{E(1-\nu)}{(1+\nu)(1-2\nu)}.
\label{equation:D}
\end{equation}

Equation (\ref{equation:garlekin2}) is the main general result in this
section. It needs to be complemented by boundary and initial 
conditions for displacements and pressure. 
It is not difficult to see from Eqs.(\ref{equation:garlekin1}) and 
(\ref{equation:garlekin2}) that if $curl\;u \equiv 0$ the displacements
are characterized by a vector-diffusion equation. However, the general
case with non-vanishing curl is much more complicated (it is described
by an integro-differential equation in ${\bf u}$).
 
\subsubsection{Dispersion relation}\label{subsubsection:dispersion_relation}
While theoretical simplifications within models are usually appreciated 
a remaining question is whether the simplified model assumptions do 
lead to a consistent physical picture concerning certain criteria. One such 
criterion is stability. For example, is it obvious that the model equations 
are stable if one disregards inertial forces for the elastic continuum phase?
This problem is most convenient considered investigating the dispersion 
relation $\omega ({\bf k})$ obtained from an expansion after eigenfunctions.
For brevity we only present the basic result in the following.

Assuming certain mathematical integrability conditions on displacements
${\bf u}({\bf r}, t)$ and pressure $p({\bf r}, t)$ both fields can 
be represented as Fourier-Integrals,
\begin{equation}
{\bf u}({\bf r}, t)=\int{d{\bf k}}\;{\bf \tilde u}({\bf k})\;
e^{i(\omega t - {\bf k}{\bf x})},
\label{equation:ansatz_fourier_u}
\end{equation}
and
\begin{equation}
p({\bf r}, t)=\int{d{\bf k}}\;\tilde p({\bf k})\;
e^{i(\omega t - {\bf k}{\bf x})}.
\label{equation:ansatz_fourier_p}
\end{equation}
Inserting this two representations into the basic 
Eqs.(\ref{equation:lame}) and (\ref{equation:flow}) one obtains 
the dispersion relation,
\begin{equation}
\omega ({\bf k})=i\;D\;{\bf k}^2,
\label{equation:dispersion}
\end{equation}
with $D$ being already defined in Eq.(\ref{equation:D})
The dispersion relation has the form of a diffusion only 
problem, i.e., all solutions are damped in time and therefore the equations 
are stable. The damping is controlled by a constant of diffusion $D$ in 
a rough estimate of order $10^{-1} \frac{m^2}{s}$ which is {\em orders} 
of magnitude larger than for molecular diffusion (we estimated $D$ from 
$\kappa \approx 10^{-15}\;m^2$, $E\approx 10^{11}\; Pa $, 
$\mu \approx 10^{-3}\;Pa\;s$ and $\nu=0.25$).

\section{The point source \\
hydraulic loading) 
}\label{section:point_source}
In the following we consider a particular simple case in three 
dimension: at time $t_0$ the material is loaded by a pressure peak
$\delta(t)\delta({\bf r})$.
Dimensionless quantities for the length scale, stress 
and time scale will be useful (denoted by a star);
\begin{equation}
\label{equation:units}
{\bf r}^{*}= \frac{{\bf r}}{\kappa^{1/2}};\;\;
p^{*}= \frac{p}{G};\;\;
t^{*}= t \frac{G}{\mu},
\end{equation}
with $\kappa$, $2G=E/(1+\nu)$ and $\mu$ being the 
permeability, elastic modulus and fluid viscosity respectively.

\subsection{Pressures and Displacements}
\label{subsection:pressures_and_displacements}
\begin{figure}[htb]
\centerline{\psfig{file=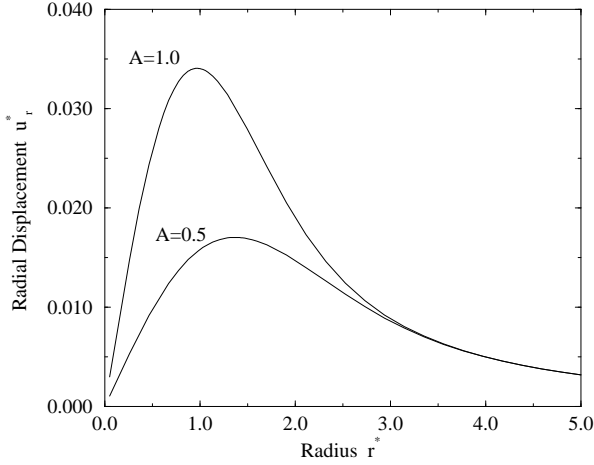,width=0.5\textwidth}}
\vskip 5mm
\caption{Radial displacement according to 
Eq.(\ref{equation:ur}) for two different 
times $A =1.0$ and $A =0.5$.
}
\label{fig:ur}
\end{figure}

Because the chosen symmetry of boundary conditions guarantuees
the existence of an irrotational solution Lam\'e's equation 
Eq.(\ref{equation:lame}) simplifies considerable,
\begin{equation}
\label{equation:curlfree}
div^{*}\;{\bf u}^{*}= {a^{*}}^2  (p^* - p_0^*),
\end{equation}
where $p_0^*$ denotes the equilibrium pressure and 
$2{a^{*}}^2 = (1-2\nu)/(1-\nu)$  a material constant.
Note that the solution for ${\bf u}^*$ can be obtained by direct 
integration of Eq.(\ref{equation:curlfree}), 
\begin{equation}
\label{equation:generic_solution}
{\bf u}^*({\bf r}^*)= -\frac{{a^*}^2}{4\pi}\;grad^*
\int_{V'}\;{dV^*}'\;
\frac{p^*({\bf {r^*}'})-p_0^*}{\vert {\bf r^*}-{\bf {r^*}'}\vert}.
\end{equation}
Equation (\ref{equation:curlfree}) can be inserted into 
the pressure equation (\ref{equation:flow}) giving an (inhomogeneous)
heat conductance equation,
\begin{equation}
\label{equation:heat_dirac}
\Delta^*\;p^*\; - {a^{*}}^2 \frac{\partial p^*}{\partial t^*} =
-\delta(t^*)\delta({\bf r^*}), 
\end{equation}
with three dimensional solution (Green's function),
\begin{equation}
\label{equation:heat_dirac_solution}
p^*({\bf r^*},\; t^*)-p_0^* = \frac{1}{{a^*}^2}
(A/\pi)^{3/2} e^{-A {r^*}^2},
\end{equation}
where,
\begin{equation}
\label{equation:A}
A = \frac{{a^*}^2}{4 t^*}=
\frac{(1-2\nu)\mu }{8(1-\nu)Gt}.
\end{equation}
Thus incompressible flow in a porous medium whose displacement field 
is irrotational corresponds to a heat conductance problem in 
a fiktive non-porous medium. 
One can insert the pressure field
Eq.(\ref{equation:heat_dirac_solution})
into Eq.(\ref{equation:generic_solution}) and evaluate the integral,
\begin{equation}
\label{equation:ur}
{\bf u}^*({\bf r}^*)= \frac{1}{2\pi^{3/2}} \Bigl(
\frac{1}{{{\bf r}^*}^2} 
\int_0^{\sqrt{A}\vert{\bf r}^*\vert}e^{-z^2}\,dz -
\frac{\sqrt{A}}{\vert{\bf r}^*\vert} e^{-A {{\bf r}^*}^2} \Bigr)
{\bf e}_r .
\end{equation}
In Fig.\ref{fig:ur} we show the dimensionless radial displacements, 
$u^*_r$, for two different times.
The displacements being zero at the symmetry center raise
to a maximum value and drop for larger radii towards zero again.
As can be seen from Eq.(\ref{equation:ur}) do the
displacements follow a scaling form. 
The scaling variable $\delta = \sqrt{A}{\bf r^*}$ may be 
anticipated for a diffusion-related problem.
%\begin{figure}[htb]
%\centerline{\psfig{file=urr.ps,width=0.5\textwidth}}
%\vskip 5mm
%\caption{Radial deformation $u^*_{rr}=u^*_{r,r}$ at  two different times, 
%$A=1.0$ (early times) and $A=0.5$ (later times), 
%comp.~Eq.(\ref{equation:A}). Note the tensional stresses for 
%small radii $r^*$ due to the finite fluid viscosity.
%}
%\label{fig:urr}
%\end{figure}

\subsection{Deformations}
\label{subsection:deformations}
\begin{figure}[htb]
\centerline{\psfig{file=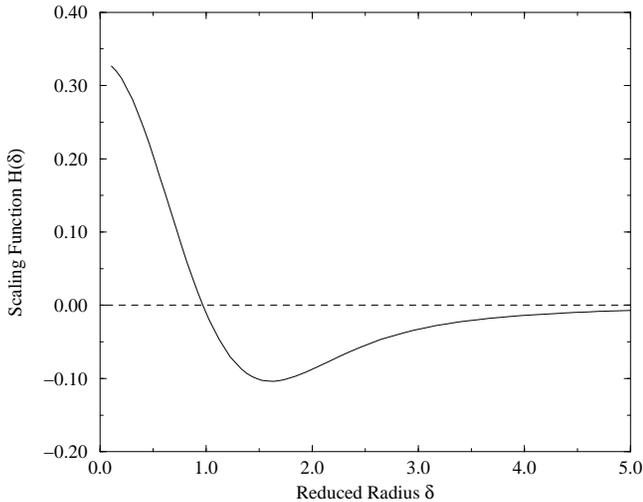,width=0.5\textwidth}}
\vskip 5mm
\caption{
Scaling function $H(\delta)$ as function of the reduced radius 
$\delta=\sqrt{A}r^*$, comp.~ Eqs.(\ref{equation:units}), 
(\ref{equation:A}) and (\ref{equation:H_delta}).
The function, being essentialy a rescaled radial deformation, shows four 
regions of interest: (a) for small radii the positive large but finite 
deformation, (b) for $\delta_0 \approx 1$ a root, (c) a negative, minimum 
deformation at $\delta_c \approx 1.6$ and (d) a decay towards zero for large 
radii.
}
\label{fig:urr}
\end{figure}

It is obvious that the deformations can be written in scaling 
form,
\begin{equation}
u^*_{rr}=\bigl( \frac{A}{\pi} \bigr)^{3/2} H(\delta),   
\label{equation:deformation_scaling}
\end{equation}
with $H(\delta)$ being a scaling function of the form,
\begin{equation}
H(\delta) = e^{-\delta^2} + \frac{e^{-\delta^2}}{\delta^2} -
	\frac{1}{\delta^3}\int_0^\delta dz\,e^{-z^2},
\label{equation:H_delta}
\end{equation}
being plotted in Fig.~\ref{fig:urr}.

Let $\delta_0$ denote the root of the forgoing equation. We find,
\begin{equation}
\delta_0 = \sqrt{A}r^*_0 \approx 1, 
\end{equation} 
within three percent error. The volumes exhibiting positive and negative 
stresses are therefore separated by a surface of radius $r^*_0$ at time 
$A$ around the center, or equivalently $r_0= 2\sqrt{Dt}$ with diffusity
$D$ defined in Eq.~(\ref{equation:D}), eg.
an order of magnitude estimate gives $r_0 \approx 20\,cm$ 
for $t=0,1\,sec$ for the neutral zone. 
Another interesting property is the asymptotic behaviour of the deformation 
for $\delta \to \infty$. The asymptotics is given as,
\begin{equation}
\lim_{\delta\to\infty} \frac{\partial u^*}{\partial r^*}=
%\frac{\partial u}{\partial r}= 
-\frac{1}{2\pi {r^*}^3}=-\frac{1}{2\pi}
\Bigl( \frac{\kappa^{1/2}}{r}\Bigr)^3 .
\label{equation:urr_infinite}
\end{equation}
It is interesting to note that neither elastic constants nor fluid properties 
enter this relation. Only the permeability enters this compressional
{\em far-field} decay.
This relation should be compared to the simple case of a cavity of radius 
$R$ bearing pressure $-p$ within an infinite non-Biot medium with 
vanishing pressure at infinity,
\begin{equation}
u_{rr}=-p\frac{(1+\nu)}{E}\Bigl(\frac{R}{r}\Bigr)^3=
-\frac{p^*}{2}\Bigl(\frac{R}{r}\Bigr)^3,
\end{equation}
when one measures the bearing pressure in units of the shear modul 
$G=\frac{E}{2(1+\nu)}$.
Both relations are essentially the same, where the {\em typical} length-scale
$\kappa^{1/2}$ in Eq.(\ref{equation:urr_infinite}) given by the solids 
permeability acts similar to the cavity-size in non-Biot elasticity.
In principle this should allow  to {\em measure} the permeability 
experimentally for $\delta \to \infty$, which means in practice for distances 
$r>> 2\sqrt{Dt}$. This observation leads us to the conclusion 
that the more complicated 'Biot elasticity' reduces for length scales 
much larger than the scale of diffusion towards convential linear elasticity.

The maxima of the tensile deformations, $\delta\to 0$, are readily calculated 
using $\lim_{\delta\to 0}H(\delta)=1/3$ \cite{my_hospital},
\begin{equation}
\lim_{\delta\to 0}\frac{\partial u^*}{\partial r^*}=
\frac{A^{3/2}}{3\pi^{3/2}}.
\end{equation}
This value does {\em not} depend on the material's permeability.

\begin{figure}[htb]
\centerline{\psfig{file=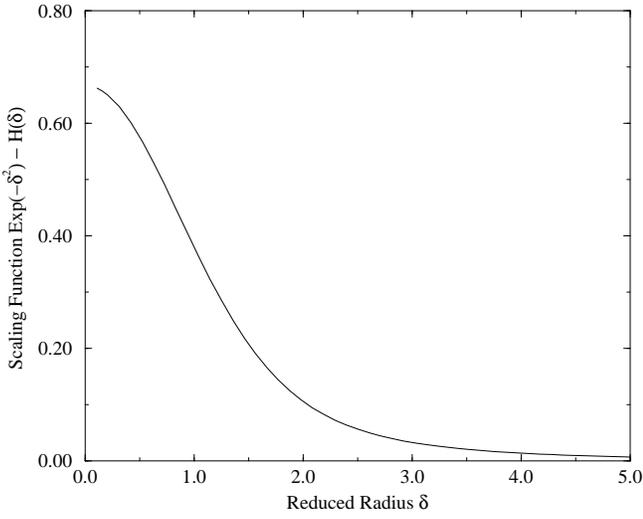,width=0.5\textwidth}}
\vskip 5mm
\caption{Polar deformation $u^*_{\Phi\Phi}=u^*_r/r^*$. 
Because the materials moves outwards from the 
injection center all deformations are positive. Note that the polar 
deformations are monotonically decreasing for increasing $\delta$ while
the radial deformations exhibit a negative minimum..
}
\label{fig:upp_delta}
\end{figure}

Before turning towards the mechanical stresses we quickly summarize the 
results for the polar deformations $u^*_{\phi\phi}$. 
Fig.~\ref{fig:upp_delta} shows the polar deformations. in a rescaled form. 
One has,
\begin{equation}
u^*_{\phi\phi}=\frac{1}{2}\Bigl( \frac{A}{\pi} \Bigr)^{3/2} 
\Bigl( e^{-\delta^2} - H(\delta) \Bigr),
\label{equation:upp_delta}
\end{equation}
with $H(\delta)$ and $\delta$ as defined in Eq.~(\ref{equation:H_delta}).
The limiting cases are,
\begin{equation}
\lim_{\delta\to 0}u^*_{\phi\phi}=\lim_{\delta\to 0}u^*_{rr},  
\end{equation}
and 
\begin{equation}
\lim_{\delta\to \infty}u^*_{\phi\phi}=-\frac{1}{2}
\lim_{\delta\to \infty}u^*_{rr}.
\end{equation}

\subsection{Stresses}
\label{subsection:stresses}
\begin{figure}[htb]
\centerline{\psfig{file=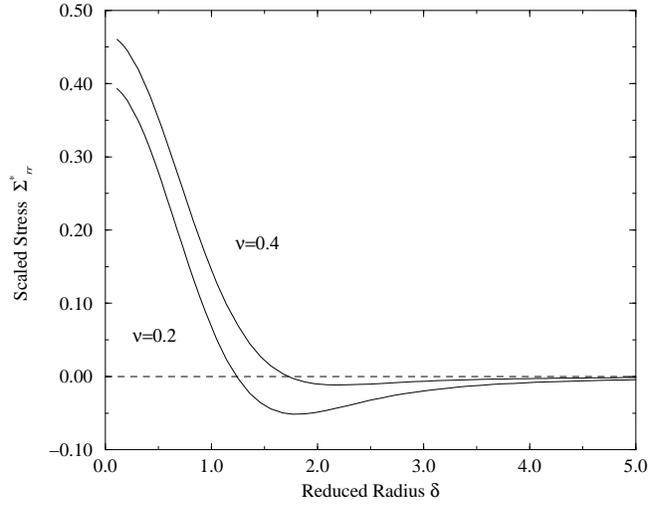,width=0.5\textwidth}}
\vskip 5mm
\caption{
Rescaled radial stress $\Sigma^*_{rr}=\frac{1-2\nu}{2}\Bigl( 
\frac{A}{\pi} \Bigr)^{-3/2}\sigma^*_{rr}$ as a function of the rescaled 
radius $\delta$ for two values of Poisson numbers.
For $\nu\to 1/2$ the scaling function decays like 
$\Sigma^*_{rr}=e^{-\delta^2}$, which is positive only.
}
\label{fig:stresses_delta}
\end{figure}

We now proceed to calculate the radial stresses $\sigma^*_{rr}$ 
which we measure in units of the shear modul $G$,
\begin{equation}
\sigma^*_{rr}= \frac{2}{(1-2\nu)}
\Bigl( (1-\nu)u^*_{rr} +2\nu u^*_{\phi\phi} \Bigr).
\label{equation:stresses}
\end{equation}
which is equivalent to
\begin{equation}
\frac{1-2\nu}{2}\Bigl( \frac{A}{\pi} \Bigr)^{-3/2}\sigma^*_{rr}
=(1-\nu)H(\delta) + \nu (e^{-\delta^2}- H(\delta) ).
\label{equation:stresses_delta}
\end{equation}
We have plotted the rescaled stresses Eq.~(\ref{equation:stresses_delta})
in Fig.~\ref{fig:stresses_delta} for Poisson numbers $\nu= 0.2$ and 
$\nu =0.4$.

The characteristic points of the rescaled stress $\Sigma^*_{rr}$ do now 
depend on $\nu$ albeit not very strongly, i.e., for lower $\nu$ one 
finds lower maximum tensile stresses, lower roots $\delta_0$ and 
higher compressive stresses.
Generally spoken do the {\em polar} deformations $u^*_{\phi\phi}$
strengthen the tensile stresses and weaken the compressive region.
Higher contraction numbers $\nu$ do stronger weight polar deformations
on radial stresses.
The two limiting cases for the radial stresses are,
\begin{equation}
\lim_{\delta\to 0}\sigma^*_{rr}= \frac{2(1+\nu)}{3(1-2\nu)}
\Bigl(\frac{A}{\pi}\Bigr)^{3/2},
\end{equation}
and 
\begin{equation}
\lim_{\delta\to \infty}\sigma^*_{rr}=-\frac{1}{\pi(1-2\nu){r^*}^3}.
\end{equation}

\section{Conclusion}
\label{section:conclusion}
We have considered a simplistic model for the fluid flow within a porous 
elastic solid. The presented description is given by linear equations.
Essential simplifications are: Vanishing inertial forces of the solid 
phase and vanishing fluid compressibility. The first assumption relieves us 
of the problem of wave propagation in Biot media. The second assumption 
allows us to construct a linear description.
We tried to obtain some general results for the linear problem.
It appears that the general integrals
for Biot's and the thermo-elastic problem 
are different even within a linear description. 
While in both cases the body force field is governed by scalar potentials,
e.g. pressure and temperature respectively, Biot's problem is a 
fully cross-coupled field problem. This changes the structure of the governing
Galerkin-Biot vector-potential, Eq.~(\ref{equation:garlekin2}), 
substantially in such a way
that {\em mixed} space-time derivatives do appear.
However, we were able to show 
that an irrotational displacement field garantuees an equivalence of 
both problems. Though irrotational displacement fields are rare in 
geometries of practical interest, we studied the point symmetric 
case in three dimensions in order to exemplify the fluids action 
on the elastic strain/stress distribution.
We quantitatively discussed the situation that arises when a pressure pulse
within a 'borhole' drives fluid through the solids pore-space. 
We found that radial tensile stresses arise in a region where non-permeable 
solids exhibit only compressive stresses. This surely is of interest 
to selected problem in soil meachanics and fracture mechanics, i.e. 
hydraulic fracturing.

\section*{Acknowledgments}
F.~T.~ would like to acknowledge financial support from
CEC under grant number ERBFMBICT950009 and thank for hospitality at NTNU 
Trondheim. 
Many thanks also go to Alex Hansen, Magnus Wangen and Thomas 
Ihle for stimulating discussions.
  
%
%       BIBLIOGRAPHY
%

\end{multicols}
\vfill\eject


\begin{thebibliography}{10}
\bibitem{my_hospital}\label{my_hospital}
We have calculated $\lim_{\delta \to 0}H(\delta)$ employing an infinite 
recursion of L'Hospital's rule, leading to a geometric series.
\end{thebibliography}
\end{document}